\documentclass{article}
\usepackage{ijcai19}

\usepackage{times}
\usepackage{soul}
\usepackage{url}
\usepackage[hidelinks]{hyperref}
\usepackage[utf8]{inputenc}
\usepackage[small]{caption}
\usepackage{graphicx}
\usepackage{amsmath}
\usepackage{booktabs}
\usepackage{bm}
\usepackage{array}
\usepackage{amsmath}
\usepackage{multirow}
\usepackage{subfigure}
\usepackage{stfloats}

\newcommand{\spara}[1]{\smallskip\noindent\textbf{#1}.}
\newcommand{\eat}[1]{}
\newcommand{\tabincell}[2]{\begin{tabular}{@{}#1@{}}#2\end{tabular}}

\title{Disparity-preserved Deep Cross-platform Association \\ for Cross-platform Video Recommendation}

\author{
Shengze Yu$^1$\and
Xin Wang$^{1}$\footnote{Corresponding Author}\and
Wenwu Zhu$^{1*}$\and
Peng Cui$^1$\And 
Jingdong Wang$^2$
\affiliations
$^1$ Department of Computer Science and Technology, Tsinghua University, China\\
$^2$ Microsoft Research
\emails
ysz0323@163.com,
\{xin\_wang, wwzhu, cuip\}@tsinghua.edu.cn,
jingdw@microsoft.com
}

\begin{document}

\maketitle

\begin{abstract}
Cross-platform recommendation aims to improve recommendation accuracy through associating information from different platforms.
Existing cross-platform recommendation approaches assume all cross-platform information to be consistent with each other and can be aligned. 
However, there remain two unsolved challenges:
i) there exist inconsistencies in cross-platform association due to platform-specific disparity, and
ii) data from distinct platforms may have different semantic granularities.
In this paper, we propose a cross-platform association model for cross-platform video recommendation, i.e., Disparity-preserved Deep Cross-platform Association (DCA), taking platform-specific disparity and granularity difference into consideration. 
The proposed DCA model employs a partially-connected multi-modal autoencoder, which is capable of explicitly capturing platform-specific information,
as well as utilizing nonlinear mapping functions to handle granularity differences. 
We then present a cross-platform video recommendation approach based on the proposed DCA model. 
Extensive experiments for our cross-platform recommendation framework on real-world dataset demonstrate that the proposed DCA model significantly 
outperform existing cross-platform recommendation methods in terms of various evaluation metrics.
\end{abstract}

\section{Introduction}
 
\eat{With the explosive growth of various multimedia content sites, watching video has become an important information source for people. However, watching video is very time consuming, let alone looking through all the tremendous amount of videos, so video recommendation is of great significance.} 

Recommender systems are playing an important role in our life. 
With the emergence of various online services, people are now getting used to engaging on different platforms simultaneously in order to meet their increasing diverse information needs \cite{chen2012more}.
The complementary information from various platforms jointly reflects user interests and preferences, providing us with a great opportunity to tackle the data sparsity problem and improve the recommendation accuracy through associating information across platforms.
\eat{Given the ultimate goal of serving people intelligently, organically and optimally transferring or associating the cross-platform information becomes significantly important \cite{jiang2016little}.}
In order to improve user experience and increase user adoption, many online service providers release new features to encourage cross-platform data associations.
For example, Google+ encourages users to share their homepages on other platforms, enabling different accounts of the same person to be linked together. 
To date, quite a number of various platforms tend to associate with each other, which provides chances for cross-platform information transfer and association analysis possible.

\eat{
Typical recommendation methods, however, are usually designed for isolated single platform~\cite{yan2015unified}. 
The limitation of these single-platform based solutions is that necessary information of new and inactive users is usually insufficient or even not available for making accurate recommendations. 
The notorious cold-start and sparsity issues have significantly hindered accurate user modeling and practical personalized services~\cite{ricci2011recommender}. 
Thus in order to improve user experience and increase user adoption, many online service providers release new features to encourage cross-platform data associations. 
For example, Google+ encourages users to share their homepages on other platforms, enabling different accounts of a same person to be linked together. 
To date, quite a number of various platforms tend to associate with each other, which makes information transfer and association analysis across different platforms possible through processing cross-platform data.
}

Existing cross-platform recommendation works try to improve their cross-platform performances mainly by investigating the ablity of cross-platform association. Yan et al. propose a user-centric topic association framework to map cross-platform data in a common latent space~\shortcite{yan2014mining}. They then introduce a predefined micro-level metric to adaptively weight data while doing the data integration~\shortcite{yan2015unified}. Man et al. propose an embedding and mapping framework, EMCDR, to represent and associate users across different platforms~\shortcite{Man2017Cross}. 
However, existing works on cross-platform video recommendation ignore the inconsistencies in cross-platform association and differences in semantic granularities, two challenging phenomena discovered in this paper.

The first challenge, inconsistency in cross-platform association, results in a reconsideration for the current assumption that 
all cross-platform information of the same user is consistent and can be aligned, 
which will be empirically validated on a real-world cross-platform dataset later.
Our discovery demonstrates that the inconsistency is mainly caused by the platform-specific disparity,
i.e., in addition to a user's inherent personal preferences, its interests shown on different platforms also contain platform-specific factors due to different focuses of various platforms.
The second challenge is that different platforms may have different semantic granularities. 
Take Twitter and YouTube as an example, tags on Twitter are mainly hot event topics, which are generally finer-grained than video categories on YouTube. 
As such, a good cross-platform video recommendation model based on the correlated information across platforms should have the ability to reveal characteristics at different semantic granularities.


To address these challenges, we propose the Disparity-preserved Deep Cross-platform Association (DCA) model for cross-platform recommendation, 
which employs a partially-connected multi-modal autoencoder to explicitly capture and preserve platform-specific disparities in latent representation. 
Besides, we also introduce nonlinear mapping function, 
which is superior to their linear competitors in terms of handling granularity difference.
To validate the advantages of our proposed DCA model, we further design a cross-platform video recommendation approach based on the proposed DCA model, 
which is capable of automatically targeting and transferring useful cross-platform information from a comprehensive semantic level. 
The contributions of this paper can be summarized as follows:

\begin{itemize}
\item We recognize the inconsistency phenomenon, which is caused by platform-specific disparity, as well as the granularity difference problem in cross-platform association.
\item We propose a novel {\bf D}isparity-preserved Deep {\bf C}ross-platform {\bf A}ssociation (DCA) model for cross-platform recommendation through employing a modified multi-modal autoencoder, which is able to handle the cross-platform inconsistency issue by preserving the platform-specific disparities and solve the granularity difference problem via nonlinear functions.
\item We present a cross-platform video recommendation approach based on the proposed DCA model to test its effectiveness compared with other approaches.
\item We conduct experiments on a real-world cross-platform dataset on Twitter and YouTube to demonstrate the superiority of our proposed DCA model over several state-of-the-art methods in terms of various evaluation metrics.
\end{itemize}

\section{Related Work}
Existing cross-platform recommendation works can be categorized in two ways. 

\spara{Categorization by what to associate}
One option is to take advantage of different platform characteristics towards collaborative applications. 
Particularly,
Qian et al. propose a generic cross-domain collaborative learning framework based on nonparametric Bayesian dictionary learning model for cross-domain data analysis~\shortcite{qian2015cross}. 
Min et al. propose a cross-platform multi-modal topic model which is capable of differentiating topics and aligning modalities~\shortcite{min2015cross}.
An alternative option is to associate in a user-centric way, which focuses on integrating multiple sources of overlapped user's activities. 
In particaular,
XPTrans~\cite{jiang2016little} optimally bridges different platforms through exploiting the information from a small number of overlapped crowds. 
Yan et al. propose an overlapped user-centric topic association framework based on latent attribute sparse coding, and prove that bridging information from different platforms in common latent space outperforms explicit matrix-oriented transfer~\shortcite{yan2014mining}. 
Man et al. propose an embedding and mapping framework, EMCDR, where user representations on different platforms are first learnt through matrix factorization and then mapped via multi-layer perceptron~\shortcite{Man2017Cross}.
Although the user-centric works above are based on different premises, they share the same core idea that all cross-platform information is consistent and should be aligned.
However, several works have pointed out the data inconsistency phenomenon in cross-platform data association, and attempted to solve this problem by data selection. 
To be concrete, Lu et al. find that selecting consistent auxiliary data is important for cross-domain collaborative filtering~\shortcite{lu2013selective}. 
They propose a novel criterion to assesses the degree of consistency, and embed it into a boosting framework to selectively transfer knowledge. 
Yan et al. divide users into three groups and introduce a predefined micro-level user-specific metric to adaptively weight data while integrating information across platforms~\shortcite{yan2015unified}. 
Our proposed DCA model associates information from different platforms in the user-centric way. 

\spara{Categorization by entire model structure}
One group of methods~\cite{Chen2012TLRec,jiang2016little,jing2014transfer,qian2015cross} build the model in a unified framework, 
where the former two works adopt matrix factorization and the latter two employ the probabilistic models. 
Another group of works~\cite{yan2014mining,yan2015unified,Man2017Cross} adopt a two-step framework. 
These works first map users from different platforms into corresponding latent spaces for representations, and then associate these representations. 
Our proposed model in this work utilizes the two-step structure in a topic-based way.

\section{Measurement and Observation}

\begin{figure}[t]
    \centering
    \subfigure[User groups clustered on Twitter]{
    \begin{minipage}{0.95\linewidth}
    \centering
    \includegraphics[width=0.95\linewidth]{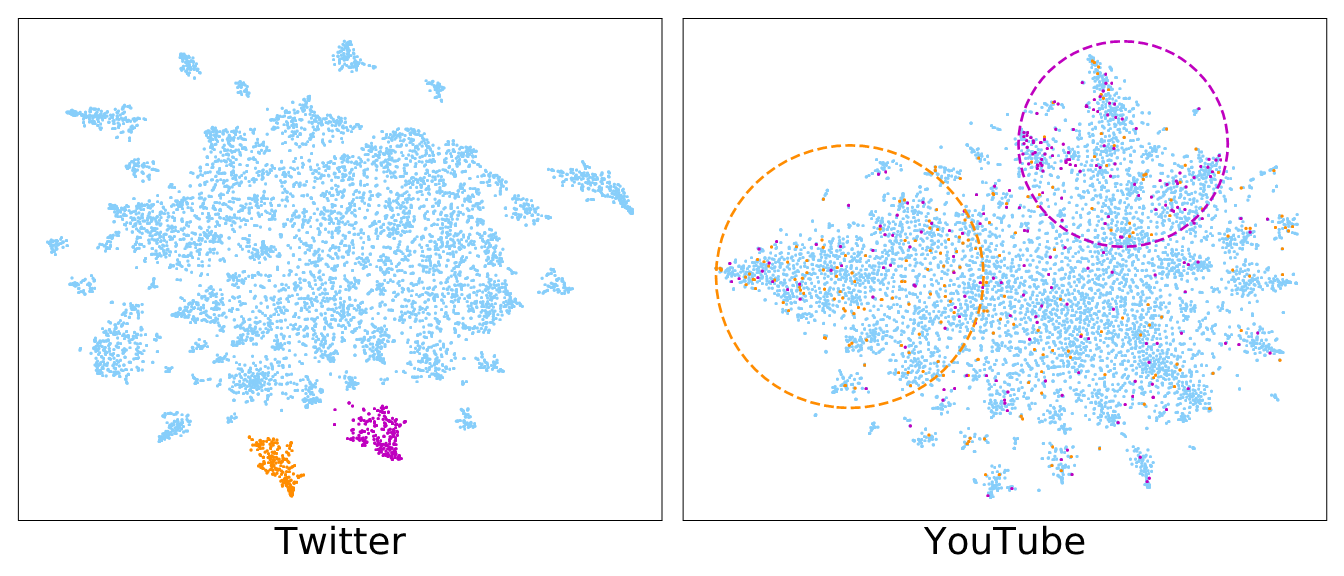}
    \label{fig:visualize_a}
    \end{minipage}
    }
    
    \vspace{-2mm}
    
    \subfigure[User groups clustered on YouTube]{
    \begin{minipage}{0.95\linewidth}
    \centering
    \includegraphics[width=0.95\linewidth]{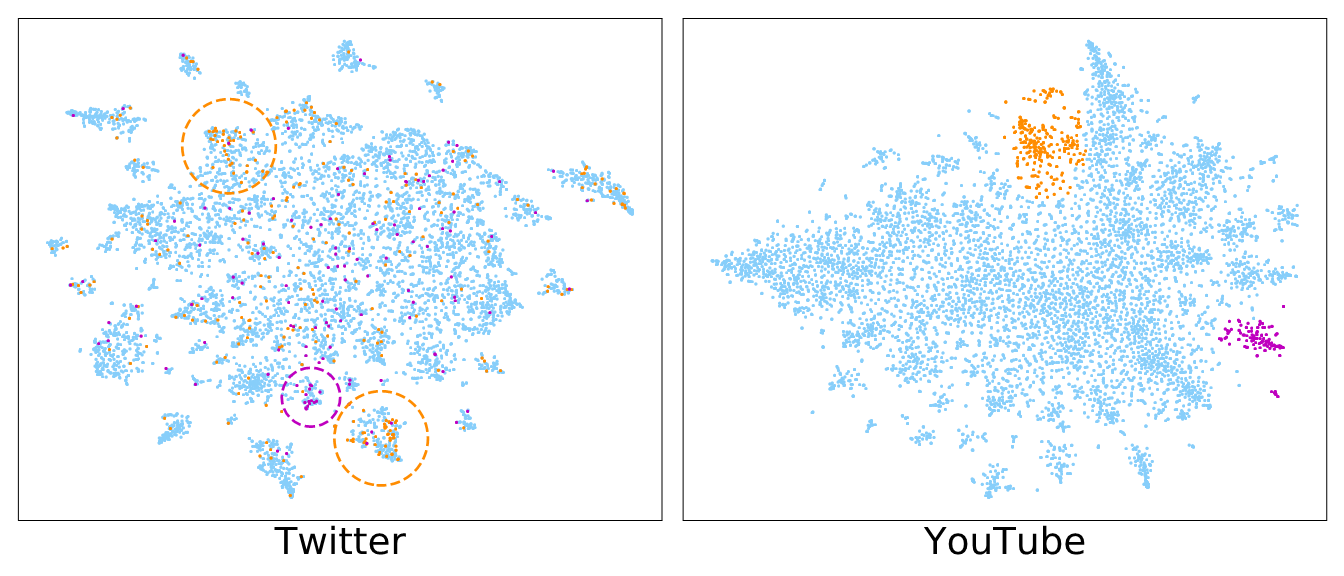}
    \label{fig:visualize_b}
    \end{minipage}
    }
    
    \vspace{-2mm}
    
    \caption{Visualization of users on Twitter and YouTube in the CASIA-crossOSN subset. Orange and purple dots represent two groups of aligned users across platforms. Concentrated areas are circled with dashed lines.}
    \label{fig:visualize}
\end{figure}

To validate the existence of the inconsistency, we first visualize users in CASIA-crossOSN, a cross-platform dataset linking user accounts between YouTube and Twitter (whose detailed description will be given in Section~\ref{sec:exp_DCA}), following three steps as follows.
1) For users who engage in both Twitter and YouTube, we extract their interest representations by Latent Dirichlet Allocation (LDA).
2) We adopt the interest representations as inputs to the visualization tool t-SNE~\cite{maaten2008visualizing} so that users are represented as two-dimensional vectors and visualized as dots --- the closer two dots are in distance, the more similar their corresponding users are in topic space. 
3) We cluster users into different groups by their Twitter (resp. YouTube) interest representations, select two groups, and mark them with orange and purple respectively in Figure~\ref{fig:visualize} such that users who share a same color are regarded as similar.

We observe from Figure~\ref{fig:visualize} that only a part of users who are similar on Twitter (resp. YouTube) still stay similar on YouTube (resp. Twitter) (circled with dashed line) and the rest actually becomes quite different. Moreover, in Figure~\ref{fig:visualize}\subref{fig:visualize_b}, orange user group clustered on YouTube splits into two subset on Twitter, indicating that two platforms may have different semantic granularities. 

\begin{table}[t]
\centering
\small
\begin{tabular}{m{1.2cm}<{\centering}|m{1.2cm}<{\centering}|m{2.1cm}<{\centering}|m{2.1cm}<{\centering}}
\hline
\multirow{3}*{Platform} & \multicolumn{3}{c}{\tabincell{c}{Average Euclidean distance\\from center of user groups}} \\ \cline{2-4}
                & Random        & Clustered on Twitter  & Clustered on YouTube  \\ \hline
Twitter			& 0.346 (1)     & 0.244 (0.705)         & 0.341 (0.985)         \\ \hline
YouTube         & 0.224 (1)     & 0.223 (0.995)         & 0.198 (0.883)         \\ \hline
\end{tabular}
\caption{Average Euclidean distance from the center of user groups clustered on Twitter (resp. YouTube) vs. random. User groups clustered on Twitter (resp. YouTube) no longer concentrate on YouTube (resp. Twitter), indicating cross-platform user interest disparity.}
\label{tab:measurement}
\end{table}

In addition to visualizing the interest representations, Table~\ref{tab:measurement} further confirms the existence of the inconsistency phenomenon through a measurement study on CASIA-crossOSN.
i) For each clustered user group on Twitter and YouTube, we calculate the average Euclidean distance from the group center to group members.
ii) We do the same thing for randomly sampled groups on both platforms and use the calculated average Euclidean distances as normalizers to obtain the ``concentration ratio'' such that a larger concentration ratio indicates a less clustered pattern. 

It can be observed from Table~\ref{tab:measurement} that when we cluster users by their interest representations on Twitter (resp. YouTube), 
the concentration ratio is small on the origin platform while becomes quite large (close to 1) on the other platform, indicating that users having similar interests on one platform tend to behave quite similarly to randomly sampled users on the other 
--- they no longer share similar interests. 

We close this section by giving a conclusion on the existence of inconsistency in cross-platform association: 
users' interests on different platforms may be diverse and inconsistent.

\section{DCA: Disparity-preserved Deep Cross-platform Association}

In this section, we employ multi-modal autoencoder to present our disparity-preserved deep cross-platform association (DCA) model.

\subsection{Problem Formulation}

We first introduce the problem of user-centric cross-platform association. Given a set of aligned users $U$, for each user $u \in U$, his or her representations on different platforms ($\bm{u}^T$ on Twitter and $\bm{u}^Y$ on YouTube in the experiment) are known. The goal is to find the association among these representations (direct mapping function or unified latent representation), so that cross-platform applications can be carried out (specifically inferring $\bm{u}^Y$ for video recommendation in the experiment).

\subsection{Disparity-preserved Deep Cross-platform Association}

Multi-modal autoencoder is an extension to basic autoencoder. It is trained to reconstruct multiple inputs from a unified hidden layer when given multiple 
modalities of the same entity, thus is able to discover correlations across modalities. It has been widely applied to various multi-modal tasks since being proposed \cite{ngiam2011multimodal,zhang2014start,wang2015deep}. Hong et al. extend the concept of ``multi-modal'' to ``multi-source'' \shortcite{hong2015multimodal}. In this paper, we treat representations of the same user on different platforms as multiple modalities of his or her unified latent representation, and modify the structure of multi-modal autoencoder to conduct disparity-preserved cross-platform association.
The detailed structure of multi-modal autoencoder is formulated as:

\begin{equation}
\small
    \bm{h} = g \left[ \left( \sum_i \mathbf{W}_1^i \bm{x}^i \right) + \bm{b}_1 \right], \quad \quad
    \bm{\hat{x}}^i = g \left( \mathbf{W}_2^i \bm{h} + \bm{b}_2^i \right),
\label{eq:MA}
\end{equation}

\noindent where $i \in \{T,Y\}$ denotes different platforms (e.g., Twitter and YouTube). $\bm{x}^i$ is the ``multi-modal'' input layer, 
specifically the representations $\bm{u}^i$ of the user $u$ on different platforms. $\bm{h}$ is the hidden layer, i.e., the derived unified user latent representation. 
$\bm{\hat{x}}^i$ is the output layer, i.e., the reconstructions of inputs. 
Weight matrices $\mathbf{W}$ and bias units $\bm{b}$ serve as parameters of the multi-modal autoencoder, 
denoted as $\theta$ together. $g(\cdot)$ is the activation function.
The loss function of multi-modal autoencoder is defined as follows:

\begin{equation}
\small
\begin{aligned}
    L(\bm{x}^i;\theta) &= \sum_i {\left\Vert\bm{\hat{x}}^i - \bm{x}^i\right\Vert}_2^2 
    +\lambda \sum_{\mathbf{W} \in \theta} {\left\Vert\mathbf{W}\right\Vert}_{\rm F}^2 + \mu {\left\Vert\bm{h}\right\Vert}_1,
\label{eq:loss_l1}
\end{aligned}
\end{equation}

\noindent where the first term is reconstruction error. The second term is regularizer to avoid overfitting. 
The third term is a sparsity constraint to encourage sparse latent representations. 
We minimize the loss function and train the parameters through backpropagation.

Multi-modal autoencoder takes representations of the same user on different platforms as inputs, mapping them to a unified hidden layer $\bm{h}$ 
and expecting accurate reconstructions. It learns mapping functions from representations on different platforms to the unified representation and then vice versa, 
which automatically completes the process of cross-platform association. Moreover, it is able to handle the granularity difference through introduction of nonlinearity.

We note that all multi-modal inputs should be given for the multi-modal autoencoder to work properly. 
However, in many cross-platform applications, we need to infer unknown data on other platforms, which is extremely difficult. 
One straightforward solution is to train several networks with a certain combination of several platforms as inputs 
and output the results for all platforms, setting every decoding weight equally. 
However, such an approach does not scale well as we will need an exponential number of models. 
Ngiam et al. propose to train bi-modal autoencoder using augmented but noisy data with additional examples that have only one single modality as inputs \shortcite{ngiam2011multimodal}. In this paper, we augment data with average values of different platforms, and still expect the network to ``reconstruct'' real representations for all modalities. To make the model become more robust to absent inputs and being capable of inferring unknown data in cross-platform association
task, we split the training data as follows: one-third of the training data includes user representations on Twitter and average user representation on YouTube, 
one-third consists of average user representation on Twitter and user representations on YouTube, the remaining one-third contains groundtruth representations.

\begin{figure}[t]
\centering
\includegraphics[width=0.98\linewidth]  {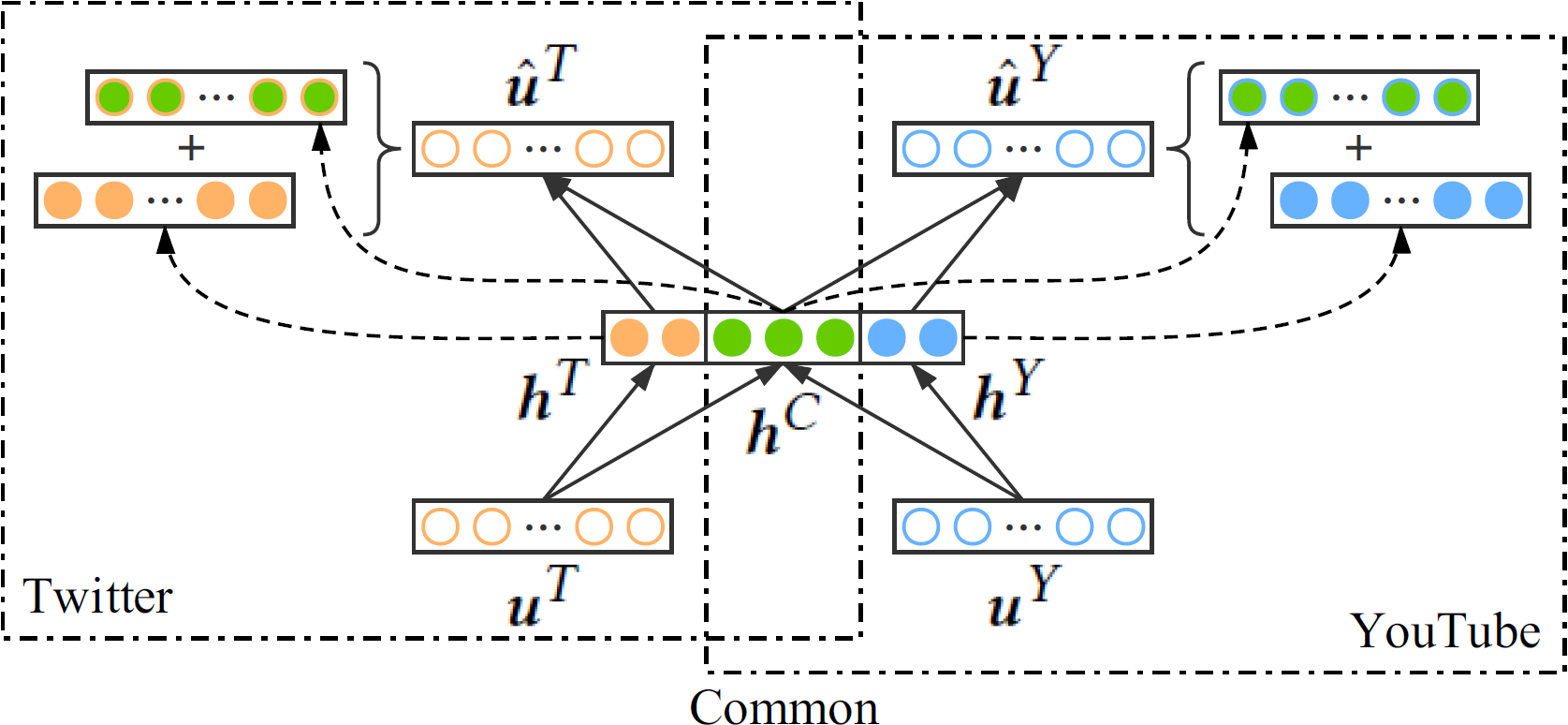}
\caption{\label{fig:DCA} Disparity-preserved Deep Cross-platform Association model. $\bm{u}^T$ and $\bm{u}^Y$ are representations of a same user on Twitter and YouTube respectively. In latent representations, $\bm{h}^T$ and $\bm{h}^Y$ are platform-specific parts preserving the disparities, and $\bm{h}^C$ is the common part associating different platforms. The estimated representations $\bm{\hat{u}}^T$ and $\bm{\hat{u}}^Y$ are derived from both common and platform-specific parts.}
\end{figure}

As aforementioned, platform-specific disparity may cause inconsistencies in cross-platform association. We need to envisage it and avoid messing up the association. To achieve this goal, we divide the hidden layer into three parts and cut off certain links, through which the shared and platform-specific parts are explicitly captured and preserved. Specifically, the hidden layer is divided into $\bm{h} = [\bm{h}^T, \bm{h}^C, \bm{h}^Y]$, where $\bm{h}^T$ and $\bm{h}^Y$ are Twitter and YouTube platform-specific parts and $\bm{h}^C$ is the common part. As is shown in Figure \ref{fig:DCA}, we assume that Twitter representations are derived from $\bm{h}^T$ and $\bm{h}^C$ without $\bm{h}^Y$ while YouTube representations are derived from $\bm{h}^Y$ and $\bm{h}^C$ without $\bm{h}^T$. The fully-connected structure is modified through cutting off links between units: from $\bm{x}^T$ to $\bm{h}^Y$, from $\bm{x}^Y$ to $\bm{h}^T$, from $\bm{h}^T$ to $\bm{\hat{x}}^Y$ and from $\bm{h}^Y$ to $\bm{\hat{x}}^T$. 
The model is able to automatically map representations of the same user on different platforms to a unified representation 
while preserving the information in platform-specific parts at the same time. 
It does not require cross-platform data to be thoroughly consistent by allowing the existences of inconsistencies. 
Thanks to the usage of shared common structure, cross-platform association becomes clearer and tighter.

In addition, most traditional linear methods associate representations on different platforms by utilizing a linear transfer matrix, 
which is formulated in Eq. \eqref{eq:LR} and Eq. \eqref{eq:LA}. 
The linear combination limits the power in modeling complex relations across different semantic granularities. 
Our modified multi-modal autoencoder naturally introduces nonlinear activation function $g(\cdot)$, 
possessing great advantages in handling the challenging granularity differences over linear methods. 
We choose Sigmoid as the activation function in this paper and other activation functions such as ReLU or tanh can also be used.

\eat{To implement the disparity-preserved model, we set a certain number of corresponding weights in $\mathbf{W}$ to zero, 
which explains the model from a different perspective: by fixing some certain parameters to zero, 
we actually add prior knowledge (i.e., existence of disparity) as constraints.}
The time complexity is $O(MNk(n_1+n_2))$ for training and $O(k(n1+n2))$ for testing. 
$M$ is the number of training iterations, $N$ is the number of users for training, $k$ is the hidden layer dimension, 
$n_1$ and $n_2$ are the input dimensions of two platforms.
Our proposed DCA model, whose structure is presented in Figure~\ref{fig:DCA}, automatically captures common information and preserves platform-specific disparities through its partially-connected structure, and meanwhile naturally adapts to granularity difference via nonlinear functions. 

\section{Cross-platform Video Recommendation}

In this section, we present the cross-platform video recommendation approach based on the proposed DCA model.

\subsection{Problem Formulation}
We first formulate the cross-platform video recommendation problem. Given a set of aligned users $U$, for each user $u \in U$, his or her behaviors on social platform are observed (i.e., tweet text $t$ in collection $T_u$ on Twitter). The goal of cross-platform video recommendation is to recommend videos for those aligned users on video platform (i.e., YouTube) through making use of the cross-platform information.

\subsection{DCA-based Cross-platform Video Recommendation}

\begin{figure}[t]
\centering
\includegraphics[width=0.98\linewidth]  {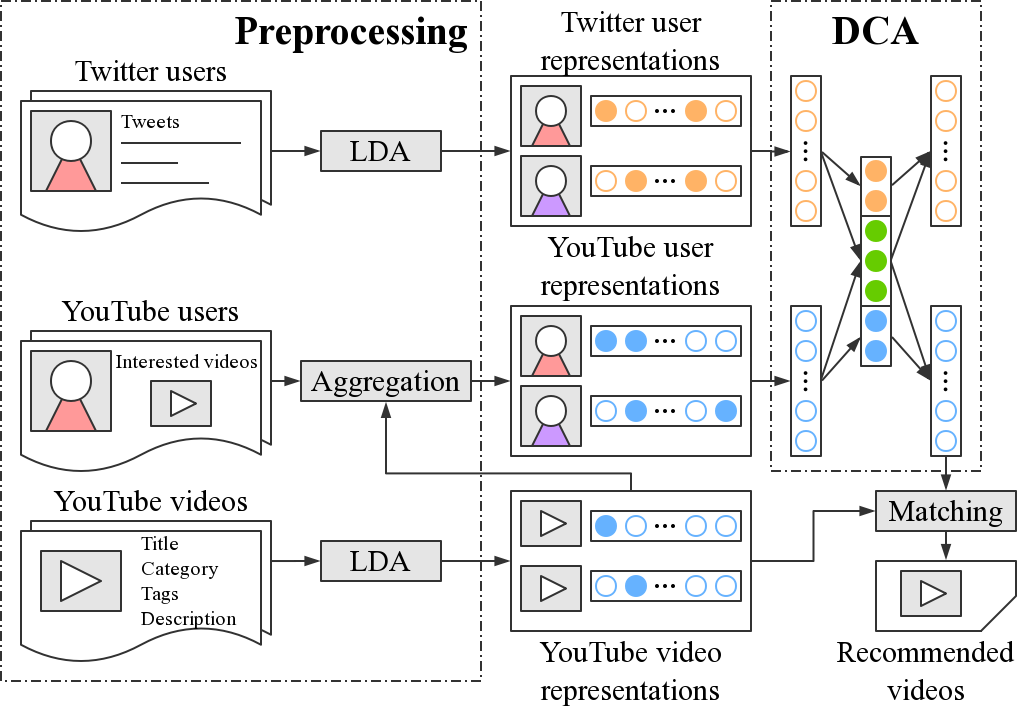}
\caption{\label{fig:framework} Cross-platform video recommendation framework based on the proposed Disparity-preserved Deep Cross-platform Association model}
\end{figure}

As is shown in Figure \ref{fig:framework}, our proposed cross-platform video recommendation method includes preprocessing and matching stages in addition to the DCA model. 
The capability of DCA model in tackling the cross-platform inconsistency issue through explicitly capturing shared information and preserving platform disparities enables the cross-platform recommendation method to automatically concentrate on useful information and discard useless one. 
Moreover, nonlinear mapping functions in our proposed approach enhance the method's ability to conduct association across platforms at a deeper and more comprehensive semantic level.

In order to apply our proposed DCA model for the associations between Twitter and YouTube, 
we first obtain user representations by Latent Dirichlet Allocation (LDA) \cite{blei2003latent}. 
On Twitter, we regard each user's tweeting data together as a document in Latent Dirichlet Allocation (LDA). 
Then each user $u \in U$ can be represented by a topic distribution $\bm{u}^T$. 
On YouTube, as we focus on the relevance between videos and users in the recommendation problem, 
we extract representations for both users and videos by two steps: 
(1) treat the surrounding textual information (title, category, tags and description) of each video together as a document 
and to get video representation $\bm{v}$ through online LDA \cite{hoffman2010online}, 
(2) take the average $\bm{v}$ of the interacted videos consumed by each user to get $\bm{u}^Y$. 
We remark that $\bm{v}$ and $\bm{u}^Y$ are in the same YouTube topic space, which enables similarity matching for recommendation.

Then the representation pairs of the same user as well as the 
augmented examples are fed to the DCA model to learn cross-platform association. 
Because of the input augmenting, the trained model is able to predict user representations on YouTube when only information from Twitter is available. 
By setting the input from YouTube to its average value, the corresponding output can be used to predict user representations $\hat{\bm{u}}^Y$.

\eat{Recommendation is finally conducted by matching with candidate videos with high Euclidean similarities in YouTube topic space.
Euclidean similarity is calculated as the properness measure:}
We finally adopt Euclidean similarities between the representations of the predicted user and candidate videos as the properness measure.
The recommendation results can be obtained according to the rank of candidate videos.

\section{Experiments}
\label{sec:exp_DCA}
In this section, we carry out extensive experiments on a real-world cross-platform dataset and compare our proposed method
with several state-of-the-art algorithms to show the advantages of the proposed approach.
\subsection{Dataset}

CASIA-crossOSN\footnote{CASIA-crossOSN dataset: \url{http://www.nlpr.ia.ac.cn/mmc/homepage/myan/dataset.html}} is a cross-network user dataset with account linkages between YouTube and Twitter, created by Institute of Automation, Chinese Academy of Sciences. It contains 11,687 aligned users across platforms and 2,280,129 YouTube videos. On YouTube, three kinds of user behaviors (i.e., uploading, favoring and adding to playlist) as well as videos' rich metadata are collected. 
On Twitter, only users' representations extracted from their tweets via standard topic modeling (LDA) process \cite{blei2003latent} are provided because of privacy concerns. We filter aligned users and YouTube videos by keeping users who interacted with at least 3 videos and videos which were consumed by at least 3 users. We finally obtain a subset containing 7,338 users and 34,144 videos, and randomly select 80\% users for training, and the rest 20\% for testing.

\subsection{Experimental Settings}

On Twitter, CASIA-crossOSN offers a 60-dimensional topical distribution for each user. On YouTube, we resort to perplexity to determine the number of topics. We select 80, which leads to the smallest perplexity on a 5\% held-out YouTube video subset, in our experiments.


To evaluate the effectiveness of the proposed DCA model, we implement the following baselines:

\vspace{-1mm}
\subsubsection{Linear Regression-based Association (LR)}
LR treats cross-platform association as a linear transfer problem, and pursues an explicit transfer matrix based on regression. The objective function is:

\vspace{-1mm}
\begin{equation}
\small
    \min_{\mathbf{W}} {\left\Vert \mathbf{W} \mathbf{U}^T - \mathbf{U}^Y \right\Vert}_{\rm F}^2 + \lambda {\left\Vert \mathbf{W} \right\Vert}_2,
\label{eq:LR}
\end{equation}


\vspace{-1mm}
\subsubsection{Latent Attribute-based Association (LA)}
Instead of pursuing hard transfer, Yan et al. introduce another method by discovering the shared latent structure behind two topic spaces \shortcite{yan2014mining}, whose objective function is:

\vspace{-1mm}
\begin{equation}
\small
\begin{aligned}
    \min_{\mathbf{D}^T, \mathbf{D}^T, \mathbf{S}} & {\left\Vert \mathbf{U}^T - \mathbf{D}^T \mathbf{S} \right\Vert}_{\rm F}^2 + {\left\Vert \mathbf{U}^Y - \mathbf{D}^Y \mathbf{S} \right\Vert}_{\rm F}^2 + \lambda {\left\Vert \mathbf{S} \right\Vert}_1, \\ s.t. & {\left\Vert \bm{d}_i^Y \right\Vert}_2^2 \le 1, {\left\Vert \bm{d}_j^T \right\Vert}_2^2 \le 1, \quad \forall i, j,
\label{eq:LA}
\end{aligned}
\end{equation}

The problem can be efficiently solved by sparse coding algorithm \cite{lee2007efficient} after a few transformations.

\vspace{-1mm}
\subsubsection{MLP-based Nonlinear Mapping (MLP)}
Man et al. employ MLP to cope with the latent space matching problem \shortcite{Man2017Cross}. MLP is a flexible nonlinear transformation. The optimization problem can be formulated as:

\vspace{-1mm}
\begin{equation}
\small
    \min_{\theta} \sum_{\bm{u} \in \mathbf{U}} {\left\Vert f_{mlp} (\bm{u}^T;\theta) - \bm{u}^Y\right\Vert}_2^2,
\label{eq:MLP}
\end{equation}

To train the above neural network-based model and our DCA model, we adopt Adam optimizer to minimize the loss functions.

\subsection{Evaluation of Cross-platform Association}

As for the proposed model in this paper, multi-modal autoencoder with $\mathcal{L}1$-norm but without disparity-preserved structure (MA) and the DCA model are examined.

We perform evaluations in two scenarios: (1) Infer YouTube from Twitter. Given users with their Twitter representation $\bm{u}^T$, to estimate their YouTube representation $\bm{u}^Y$. (2) Infer Twitter from YouTube in turn. As is discussed before, by setting the unknown input to zeros, we can get the corresponding output as the predicted representation $\bm{\hat{u}}$. In both scenarios, we utilize Mean Absolute Error (MAE) and Root Mean Square Error (RMSE) as evaluation metric:

\begin{equation}
\small
\begin{aligned}
    {\rm MAE}^i &= \frac{1}{\left\vert U_{test} \right\vert} \sum_{U_{test}} \frac{1}{K^i} \left\Vert \bm{\hat{u}}^i - \bm{u}^i \right\Vert_1, \\
    {\rm RMSE}^i &= \frac{1}{\left\vert U_{test} \right\vert} \sum_{U_{test}} \sqrt{ \frac{1}{K^i} \left\Vert \bm{\hat{u}}^i - \bm{u}^i \right\Vert_{\rm F}^2 }
\label{eq:MAE}
\end{aligned}
\end{equation}

\noindent where $i \in \{T,Y\}$ denotes platform, and $K^i$ is the dimension of representation. $u^i$ and $\hat{u}^i$ are the real and predicted user representation.

For model parameters, in LR model as Eq. \ref{eq:LR}, $\lambda$ is selected by line search. In LA model as Eq. \ref{eq:LA}, $\lambda$ and the number of latent attributes $m$ are selected by two-dimensional grid search. In the DCA model as Eq. \ref{eq:loss_l1}, parameters consist of $\lambda$, $\mu$ and $m$, which further divides into $[m^T,m^C,m^Y]$. We first tune the MA model (equivalent to set $m^T=m^Y=0, m=m^C$ in the DCA model) by multi-dimensional grid search, and get an optimal setting for $\lambda$, $\mu$ and $m$. We then carefully fine-tune $[m^T,m^C,m^Y]$ by a combined line-search strategy. Specifically for all randomly initialized autoencoder model, we conduct 6 experiments and take the average result.

\begin{figure}[t]
\small
\centering
\begin{tabular}{cc}
\hspace{-3mm}\includegraphics[width=0.5\linewidth]{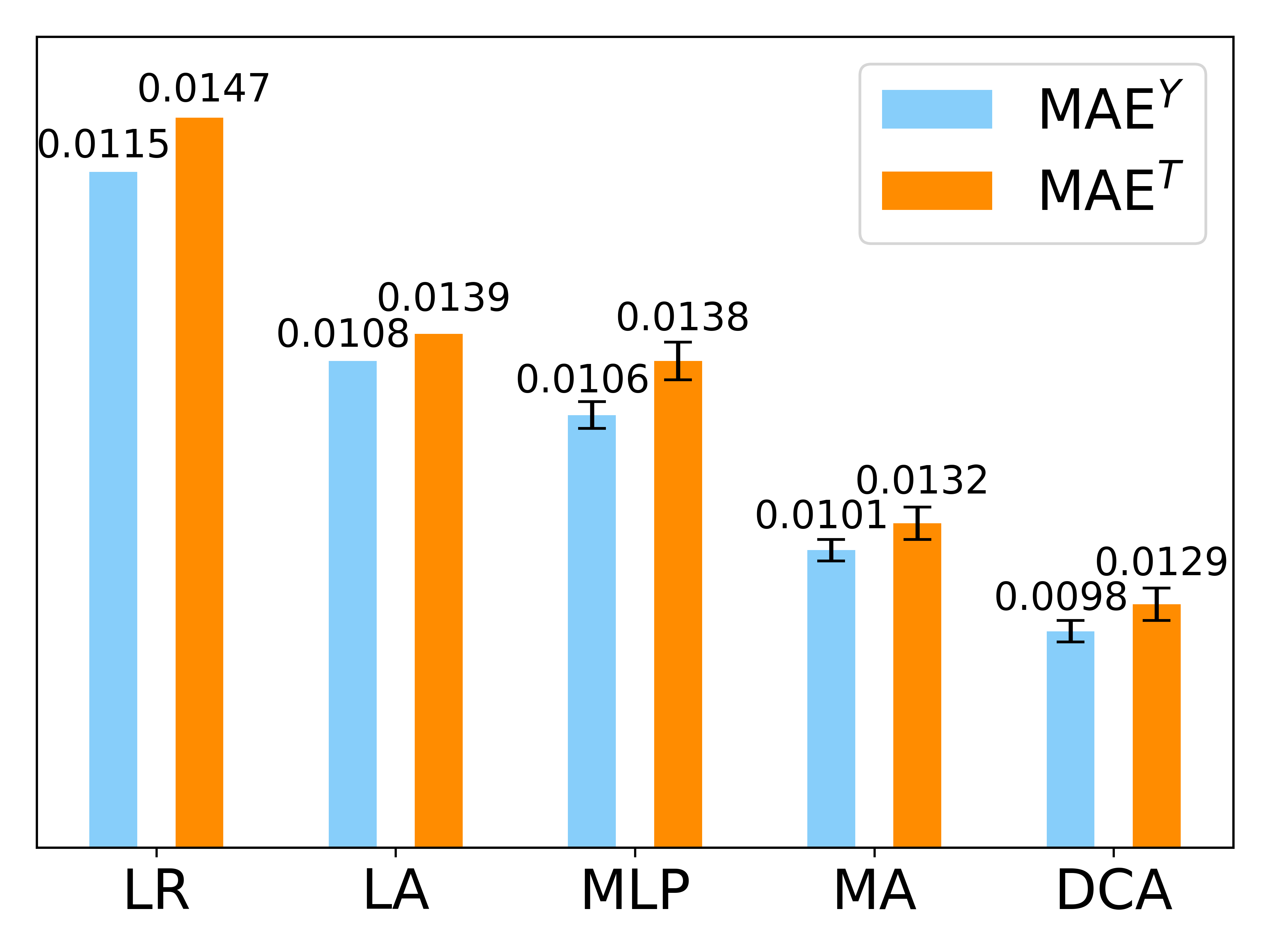} &
\hspace{-3mm}\includegraphics[width=0.5\linewidth]{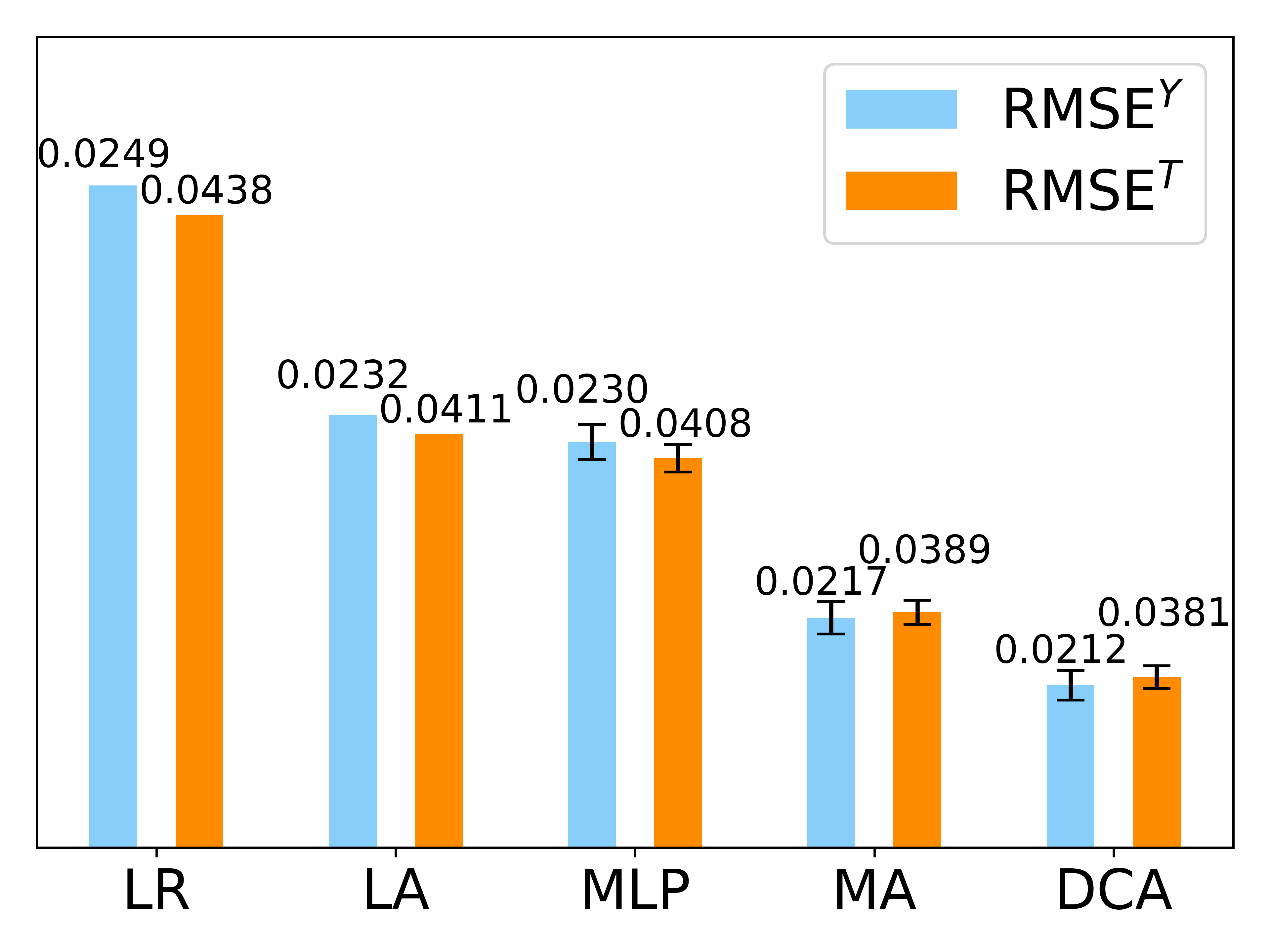} \\
\end{tabular}

\caption{${\rm MAE}^T$, ${\rm MAE}^Y$ and ${\rm RMSE}^T$, ${\rm RMSE}^Y$ of the examined models in two testing scenarios. The results show that both disparity preserving and nonlinearity contribute to better cross-platform association performance.}

\label{fig:MAE_RMSE}
\end{figure}

Performances are shown in Figure \ref{fig:MAE_RMSE} \eat{and optimal parameter settings are listed in Table \ref{tab:parameter}}. Several observations can be made: (1) LA and MLP achieve better result than LR, indicating that the shared latent structure and non-linear mapping function both benefit the association. (2) MA outperforms LA and MLP through combining the two strategies. (3) Compared with MA, DCA further improves the performance, demonstrating the contribution of the disparity-preserved structure. We also notice that compared with non-linear models (MLP, MA and DCA), the optimal parameters of the linear models vary greatly between two scenarios. We think it is the granularity differences that cause the parameter gaps. Non-linear mappings can naturally tackle this issue, and have more advantages over linear ones. 
In total, the proposed DCA model outperforms LA by reducing relatively 9.3\% on ${\rm MAE}^Y$, 8.6\% on ${\rm RMSE}^Y$, 7.2\% on ${\rm MAE}^T$ and 7.3\% on ${\rm RMSE}^T$, and outperforms MLP by reducing relatively 7.5\% on ${\rm MAE}^Y$, 7.8\% on ${\rm RMSE}^Y$, 6.5\% on ${\rm MAE}^T$ and 6.6\% on ${\rm RMSE}^T$.
It greatly improves the performance by preserving the disparities and introducing nonlinearity, validating that envisaging the inconsistency caused by platform-specific disparity and granularity difference jointly contributes to better cross-platform association performance.

\subsection{Evaluation of Cross-platform Video Recommendation}

Then we evaluate the proposed DCA-based cross-platform video recommendation method for new YouTube users, following the above ``infer YouTube from Twitter'' scenario. 

For each test user $u$, we randomly select as many as groundtruth interacted videos as additional candidates. We perform a top-$k$ recommendation task: to recommend the top $k$ YouTube videos with the highest topic-based Euclidean similarity, and adopt top-$k$ precision, recall and F-score as evaluation metrics \cite{herlocker2004evaluating}. The evaluation metrics are calculated by examining whether the recommended videos are included in $u$'s groundtruth video set $V_u$. Final results are averaged over all test users.

\begin{figure}[t]
    \centering
    \includegraphics[width=0.4\textwidth]{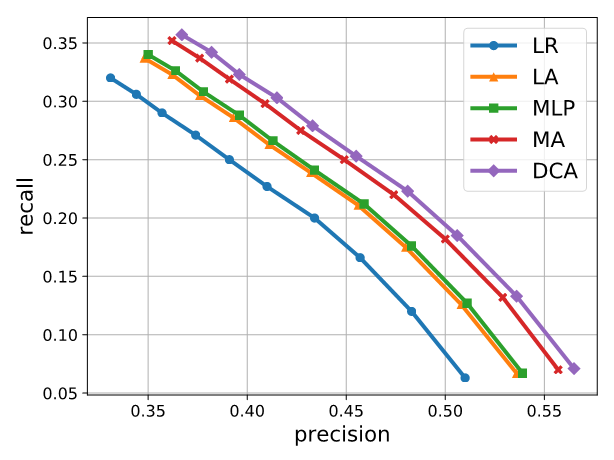}
    \caption{Precision@$k$ vs. Recall@$k$, $k$ from 1 to 10 of examined methods.}
    \label{fig:precision_recall}
\end{figure}

As is shown in Figure~\ref{fig:precision_recall} and Table \ref{tab:recommend_result}, the proposed DCA-based cross-platform video recommendation method outperforms LA-based method by enhancing relatively 6.1\% on F-score@10, and outperforms MLP-based method by enhancing relatively 5.3\% on F-score@10. The proposed DCA-based method improves over existing cross-platform association-based recommendation approaches.

\begin{table}[t]
\small
\centering
\begin{tabular}{c|c|c|c}
\hline
Model               & Precision             & Recall                & F-score           \\ \hline
LR   				& 0.331                 & 0.320                 & 0.268             \\ \hline
LA                  & 0.348                 & 0.337                 & 0.279             \\ \hline
MLP                 & 0.350                 & 0.340                 & 0.281             \\ \hline
MA                  & 0.362                 & 0.352                 & 0.290             \\ \hline
DCA                 & 0.367                 & 0.357                 & 0.296             \\ \hline
\end{tabular}
\caption{Top-10 Precision, Recall and F-score of examined methods. The results show that the proposed DCA-based cross-platform video recommendation method outperforms other approaches.}
\label{tab:recommend_result}
\end{table}

\section{Conclusion}

In this paper, we discover the existence of inconsistency in cross-platform recommendation. 
We propose the DCA model, which tackles the inconsistency issue, 
as well as the granularity difference problem. 
We further present a cross-platform video recommendation method based on the proposed DCA model.
Extensive experiments demonstrate the superiority of the DCA model and 
the DCA-based cross-platform recommendation approach over several state-of-the-art methods.

\section*{Acknowledgments}

This research is supported by National Program on Key Basic Research 
Project No. 2015CB352300, National Natural Science Foundation of China Major Project No. U1611461, China Postdoctoral Science Foundation No. BX201700136 and Shenzhen Nanshan District Ling-Hang Team Grant under No.LHTD20170005.

\clearpage
\bibliographystyle{named}
\bibliography{bibliography}

\begin{thebibliography}{}

\bibitem[\protect\citeauthoryear{Blei \bgroup \em et al.\egroup
  }{2003}]{blei2003latent}
David~M Blei, Andrew~Y Ng, and Michael~I Jordan.
\newblock Latent dirichlet allocation.
\newblock {\em Journal of machine Learning research}, 3(Jan):993--1022, 2003.

\bibitem[\protect\citeauthoryear{Chen \bgroup \em et al.\egroup
  }{2012a}]{Chen2012TLRec}
Leihui Chen, Jianbing Zheng, Ming Gao, Aoying Zhou, Wei Zeng, and Hui Chen.
\newblock Tlrec:transfer learning for cross-domain recommendation.
\newblock {\em Computer Science}, pages 167--172, 2012.

\bibitem[\protect\citeauthoryear{Chen \bgroup \em et al.\egroup
  }{2012b}]{chen2012more}
Terence Chen, Mohamed~Ali Kaafar, Arik Friedman, and Roksana Boreli.
\newblock Is more always merrier?: a deep dive into online social footprints.
\newblock In {\em Proceedings of the 2012 ACM workshop on Workshop on online
  social networks}, pages 67--72. ACM, 2012.

\bibitem[\protect\citeauthoryear{Herlocker \bgroup \em et al.\egroup
  }{2004}]{herlocker2004evaluating}
Jonathan~L Herlocker, Joseph~A Konstan, Loren~G Terveen, and John~T Riedl.
\newblock Evaluating collaborative filtering recommender systems.
\newblock {\em ACM Transactions on Information Systems (TOIS)}, 22(1):5--53,
  2004.

\bibitem[\protect\citeauthoryear{Hoffman \bgroup \em et al.\egroup
  }{2010}]{hoffman2010online}
Matthew Hoffman, Francis~R Bach, and David~M Blei.
\newblock Online learning for latent dirichlet allocation.
\newblock In {\em advances in neural information processing systems}, pages
  856--864, 2010.

\bibitem[\protect\citeauthoryear{Hong \bgroup \em et al.\egroup
  }{2015}]{hong2015multimodal}
Chaoqun Hong, Jun Yu, Jian Wan, Dacheng Tao, and Meng Wang.
\newblock Multimodal deep autoencoder for human pose recovery.
\newblock {\em IEEE Transactions on Image Processing}, 24(12):5659--5670, 2015.

\bibitem[\protect\citeauthoryear{Jiang \bgroup \em et al.\egroup
  }{2016}]{jiang2016little}
Meng Jiang, Peng Cui, Nicholas~Jing Yuan, Xing Xie, and Shiqiang Yang.
\newblock Little is much: Bridging cross-platform behaviors through overlapped
  crowds.
\newblock In {\em AAAI}, pages 13--19, 2016.

\bibitem[\protect\citeauthoryear{Jing \bgroup \em et al.\egroup
  }{2014}]{jing2014transfer}
How Jing, An-Chun Liang, Shou-De Lin, and Yu~Tsao.
\newblock A transfer probabilistic collective factorization model to handle
  sparse data in collaborative filtering.
\newblock In {\em Data Mining (ICDM), 2014 IEEE International Conference on},
  pages 250--259. IEEE, 2014.

\bibitem[\protect\citeauthoryear{Lee \bgroup \em et al.\egroup
  }{2007}]{lee2007efficient}
Honglak Lee, Alexis Battle, Rajat Raina, and Andrew~Y Ng.
\newblock Efficient sparse coding algorithms.
\newblock In {\em Advances in neural information processing systems}, pages
  801--808, 2007.

\bibitem[\protect\citeauthoryear{Lu \bgroup \em et al.\egroup
  }{2013}]{lu2013selective}
Zhongqi Lu, Erheng Zhong, Lili Zhao, Evan~Wei Xiang, Weike Pan, and Qiang Yang.
\newblock Selective transfer learning for cross domain recommendation.
\newblock In {\em Proceedings of the 2013 SIAM International Conference on Data
  Mining}, pages 641--649. SIAM, 2013.

\bibitem[\protect\citeauthoryear{Maaten and
  Hinton}{2008}]{maaten2008visualizing}
Laurens van~der Maaten and Geoffrey Hinton.
\newblock Visualizing data using t-sne.
\newblock {\em Journal of Machine Learning Research}, 9(Nov):2579--2605, 2008.

\bibitem[\protect\citeauthoryear{Man \bgroup \em et al.\egroup
  }{2017}]{Man2017Cross}
Tong Man, Huawei Shen, Xiaolong Jin, and Xueqi Cheng.
\newblock Cross-domain recommendation: An embedding and mapping approach.
\newblock In {\em Twenty-Sixth International Joint Conference on Artificial
  Intelligence}, pages 2464--2470, 2017.

\bibitem[\protect\citeauthoryear{Min \bgroup \em et al.\egroup
  }{2015}]{min2015cross}
Weiqing Min, Bing~Kun Bao, Changsheng Xu, and M.~Shamim Hossain.
\newblock Cross-platform multi-modal topic modeling for personalized
  inter-platform recommendation.
\newblock {\em IEEE Transactions on Multimedia}, 17(10):1787--1801, 2015.

\bibitem[\protect\citeauthoryear{Ngiam \bgroup \em et al.\egroup
  }{2011}]{ngiam2011multimodal}
Jiquan Ngiam, Aditya Khosla, Mingyu Kim, Juhan Nam, Honglak Lee, and Andrew~Y
  Ng.
\newblock Multimodal deep learning.
\newblock In {\em Proceedings of the 28th international conference on machine
  learning (ICML-11)}, pages 689--696, 2011.

\bibitem[\protect\citeauthoryear{Qian \bgroup \em et al.\egroup
  }{2015}]{qian2015cross}
Shengsheng Qian, Tianzhu Zhang, Richang Hong, and Changsheng Xu.
\newblock Cross-domain collaborative learning in social multimedia.
\newblock In {\em Proceedings of the 23rd ACM international conference on
  Multimedia}, pages 99--108. ACM, 2015.

\bibitem[\protect\citeauthoryear{Wang \bgroup \em et al.\egroup
  }{2015}]{wang2015deep}
Daixin Wang, Peng Cui, Mingdong Ou, and Wenwu Zhu.
\newblock Deep multimodal hashing with orthogonal regularization.
\newblock In {\em IJCAI}, pages 2291--2297, 2015.

\bibitem[\protect\citeauthoryear{Yan \bgroup \em et al.\egroup
  }{2014}]{yan2014mining}
Ming Yan, Jitao Sang, and Changsheng Xu.
\newblock Mining cross-network association for youtube video promotion.
\newblock In {\em Proceedings of the 22nd ACM international conference on
  Multimedia}, pages 557--566. ACM, 2014.

\bibitem[\protect\citeauthoryear{Yan \bgroup \em et al.\egroup
  }{2015}]{yan2015unified}
Ming Yan, Jitao Sang, and Changsheng Xu.
\newblock Unified youtube video recommendation via cross-network collaboration.
\newblock In {\em Proceedings of the 5th ACM on International Conference on
  Multimedia Retrieval}, pages 19--26. ACM, 2015.

\bibitem[\protect\citeauthoryear{Zhang \bgroup \em et al.\egroup
  }{2014}]{zhang2014start}
Hanwang Zhang, Yang Yang, Huanbo Luan, Shuicheng Yang, and Tat-Seng Chua.
\newblock Start from scratch: Towards automatically identifying, modeling, and
  naming visual attributes.
\newblock In {\em Proceedings of the 22nd ACM international conference on
  Multimedia}, pages 187--196. ACM, 2014.

\end{thebibliography}

\end{document}